# Development Trends in Steganography

Elżbieta Zielińska, Wojciech Mazurczyk, Krzysztof Szczypiorski

Warsaw University of Technology, Institute of Telecommunications

Warsaw, Poland, 00-665, Nowowiejska 15/19

**Abstract.** Steganography is a general term referring to all methods for the embedding of additional secret content into some form of carrier, with the aim of concealment of the introduced alterations. The choice of the carrier is nearly unlimited, it may be an ancient piece of parchment, as well as a network protocol header. Inspired by biological phenomena, adopted by man in the ancient times, it has been developed over the ages. Present day steganographic methods are far more sophisticated than their ancient predecessors, but the main principles have remained unchanged. They typically rely on the utilization of digital media files or network protocols as a carrier, in which secret data is embedded. This paper presents the evolution of the hidden data carrier from the ancient times till the present day and pinpoints the observed development trends, with special emphasis on network steganography.

## 1. Introduction

Year 2011 has been named by mass media "the year of the hack" [23] due to numerous accounts of data security breaches in private companies and governments. The amount of stolen data is estimated in petabytes (millions of gigabytes) [3].

Large fraction of the security breaches in 2011 can be attributed to the so called Operation Shady RAT [49]. These actions were targeted at numerous institutions around the world and the inflicted damage lasted for months in many cases. The mechanism of infection was mainly by means of conning an unaware user to open a specially crafted email (phishing) and implanting a back door on victim's computer. The next step was to connect to a website and download files which only seemed to be legit HTML or JPEG files. What really cybercriminals had done, was to encode commands into pictures or crafted web pages so they were invisible to unaware third-parties, and smuggle them through firewalls into the system under attack. Then these control commands ordered victim's computer to obtain executable code from remote servers, which in turn permitted an outsider to gain access to local files on the compromised host [32]. In numerous cases, the side channel to the confidential resources remained accessible for months, thus deeming the security breach severe. The villains were so daring that they did not even put much effort into obscuring the fact that information hiding techniques were involved in the attack. One of the pictures used as a vector for control commands was the famous "Lena", a cropped picture of a Playboy model, which is the standard test image for any digital image processing or steganographic algorithm.

It is highly probable that currently we are witnessing the birth of a whole new breed of malware. It all started with the discovery of the well-known Stuxnet [18] computer worm in June 2010 which had stirred increased attention because it had been targeted specifically to affect Iranian nuclear power plants [14]. In September of 2011 a new worm, called Duqu, was found, and it seems to be closely related to Stuxnet [7]. The general characteristics of the malware's structure are the same in both cases, however, unlike its predecessor, Duqu is oriented at gathering information on the infected system. The most stunning intricacy in Duqu's functioning is its employment of special means for transferring of the obtained data to the command and control centers of malware's authors. The captured information is hidden in seemingly innocent pictures and traverses the global network as ordinary files, without raising any suspicion [21, 51].

A similar functioning mechanism was found in a new variant of the Alureon malware [1], which was also discovered in the same period.

The abovementioned facts indicate that, in the present day's digital technologies' world, it is easily imaginable that the carrier, in which secret data is embedded, did not necessarily have to be an image or web page source code, it may have been any other file type or organizational unit of data – a packet, a frame, etc., which naturally occurs in computer networks.

However, it must be emphasized that the process of embedding secret information into an innocent looking carrier is not some recent invention – it has been known and used for ages by humankind. This process is called steganography and its origins can be traced back as far as the ancient times, but its importance has not decreased since its birth.

Among steganography's applications is providing means for conducting clandestine communication. The purpose of establishing such information exchange may be varied – possible uses can fall into the category of legal actions or illicit activity. Frequently the illegal aspect of steganography is accentuated – starting from the obvious criminal communication, through information leakage from guarded systems, cyber weapon exchange, up to industrial espionage. On the other side of the spectrum lie legitimate uses, which include circumvention of web censorship and surveillance [55], computer forensics (tracing and identification) and copyright protection (watermarking images, broadcast monitoring).

Stuxnet, Duqu, Alureon and Shady RAT are merely examples of what is becoming daily routine for security experts. What should ring the alarm bells is the incorporation of steganography into the already versatile armory of rouge hackers. It can be concluded that steganography is becoming the new black among Black Hats.

The inverse of steganography – steganalysis, which concentrates on the detection of covert communication, started to surface fairly recently, which is reflected in the proportion of available software tools concerning information hiding. Programs for the embedding of data considerably outnumber those dedicated to the detection and extraction of embedded content. The largest commercial database of steganographic tools contains 1025 applications (as of February 2012) [4], while the authors of [24] mention 111 tools dedicated to steganalysis (as of year 2007).

Let us take a closer look at the evolution of this technique, with special attention directed towards the class of methods falling in the category of network steganography.

## 2.  Recent Cases of Steganography Usage

Besides the already noted cases of steganographic methods' utilization, in the last decade, one can observe intensive research effort related to steganography and its detection methods (steganalysis). This has been caused by two facts: first, industry's and business's interest in DRM (Digital Rights Management) and second, the alleged utilization of the steganographic methods by terrorists while planning the attack on USA on 11[th] of September, 2001 [29, 45]. It is claimed that the rouge organization used images to conceal instructions regarding the plot, which were then posted on publicly available websites [45]. It seems that such communication could have passed unnoticed for as long as three years [28].

Recent findings suggest that steganography is presently exploited, mainly for illicit purposes [29, 44, 50, 57]. Robin Bryant in [11] recollects the case of "Operation Twins", which culminated in 2002 with the capture of criminals associated with the "Shadowz Brotherhood", a pedophile organization responsible for distribution of child pornography with the aid of steganography.

The mushrooming incidents involving the use of information hiding had triggered an official recognition of the problem. In the 2006 Federal Report [38], steganography had been named among the major threats of the present day networks, whose significance is predicted to be increasing. One of the solutions to alleviate the risks connected with this technique is to become acquainted with the evolution of steganography and, consequentially, predict its further development. This need has been recognized by the academic world in the early 80's, when steganography started to gain popularity.

Steganographic methods have also proven to be useful tools for data exfiltration, e.g. in 2008 it was reported [2] that someone at the U.S. Department of Justice smuggled sensitive financial data out of the agency by embedding the data in several image files.

In 2010, the revealing of a Russian spy ring of the so-called 'illegals', proved that steganography can pass unnoticed for much longer. The compromised group used digital image steganography to leak classified information from USA to Moscow [44].

# 3. The Characteristics of Steganography and Its Relationship to Cryptography

Steganography, frequently interchangeably and incorrectly referred to as information hiding, is the art of embedding secret messages (steganograms) in a certain carrier, possibly to communicate them in a covert manner. The border between the two fields cannot be visibly demarcated as their definitions are elusive, and there is a lack of a coherent classification of the invented clandestine communication methods, attributing them to specific domains of steganography or information hiding. The arising misconceptions may be attributed to the recent surge of interest in steganography observed in the mass media, which had only shed light on a small fraction of the available techniques. The reports of espionage and terrorist activities emerging during the last decade [29, 44, 50, 57] have mainly promoted these information hiding techniques, which are associated with the Internet and digital image steganography.

The question remains: how to provide rules to distinguish what belongs to the spectrum of steganographic methods. This can be done by means of providing certain conditions that must be fulfilled to consider something steganography. These may be expressed as follows:

- The information undergoing such hidden transmission is embedded in a seemingly innocent carrier, serving as camouflage for the hidden content.
- The purpose of applying a steganographic technique is to communicate information in a covert manner.
- The secrecy of the communication is guaranteed primarily by the camouflaging capability of the algorithm applied to the utilized carrier, and how well the processed data blends in with the whole bulk of legitimate entities of the cover (without embedding); this is understood as the capability to withstand detection attempts, which may rely on statistical analysis of the captured traffic or perceptual analysis of a suspicious message.

The best carrier for secret messages must possess two features. Firstly, it should be popular, i.e. the usage of such carrier should not itself be considered an anomaly. Secondly, the steganogram insertion-related modifications of the carrier should not be "visible" to the third party not aware of the steganographic procedure. Thus, if the embedding of additional information causes degradation of the carrier, then their severity should be limited to a level which would not cause suspicion.

Steganography is potentially not only limited to concealing the fact that a message is being sent, and if not detected, make the sender and receiver "invisible". Apart from this, it also provides anonymity and privacy, which become understandable desires in modern societies. Obviously, the anonymity potential of steganography, while it can be considered as beneficial in the context of protecting privacy, poses a new type of threat to individuals, societies and states. The tradeoff between the benefits and threats involves many complex ethical, legal and technological issues. In this paper we only consider the latter.

Steganography is often confused with cryptography, due to their common purpose of providing confidentiality. The difference becomes visible once the etymology of these words is known. Steganography is derived from the Greek: "covered writing", whereas cryptography stands for "secret writing". While the first describes the techniques to create a hidden communication channel, the latter is a designation of ongoing overt message exchange, where the informative content is unintelligible to unauthorized parties. To summarize, it is either the method to establish communication channel that is kept 'confidential', or the message itself. Either way, the goal of protecting information from disclosure remains common for both of these techniques – it is the means that permits us to differentiate between the two. Table 1 summarizes differences between cryptography and steganography, and Table 2 summarizes the relationship between steganography and watermarking.

Table 1 Comparison of characteristics of steganography and cryptography

|  |  | **Cryptography** | **Steganography** |
|---|---|---|---|
| **Goal** |  | Obfuscate the content of communication | Hide the fact of communication |
| **Characteristics** | Secrecy | Ciphertext is illegible | Embedded information is "invisible" to an unaware observer |
|  | Security of communication | Relies on the confidentiality of the key | Relies on the confidentiality of the method of embedding |
|  | Warranty of robustness | Complexity of the ciphering algorithm | Perceptual invisibility / statistical invisibility / compliance with protocol specification |
|  | Attacks | Detection is easy / extraction is complex | Detection is complex / extraction is complex |
| **Countermeasures** | Technical | Reverse engineering | Constant monitoring and analysis of exchanged data |
|  | Legal | Cryptography export laws | Rigid device / protocol specification |

Table 2 Comparison of characteristics of steganography and watermarking

|  |  | **Watermarking** | **Steganography** |
|---|---|---|---|
| **Goal** |  | Protect the carrier | Protect secret information from disclosure |
| **Characteristics** | Secrecy | Invisibility or perceptual visibility depending on the requirements | Embedded information is "invisible" to an unaware onlooker |
|  | Type of robustness | Robustness against tampering or removal | Robustness against detection |
|  | Effect of signal processing / random errors / compression | Must not lead to the loss of the watermark | May lead to the loss of hidden data |
|  | Type of carrier | Digital files – audio, video, text or images | Any service, protocol, file, environment employing digital representation of data |

In spite of the common historical background of the stated communication protection methods, only cryptography has managed to sustain an invariably strong position. Steganography experienced its golden age in the times of Ancient Greece and Rome, to be gradually marginalized as time passed. Intuitively, any method of protection that relies on its own confidentiality to provide the secrecy of communication, should not be publicized. Therefore, very few accounts proving contemporary exploitation of steganography can be found, which does not point to the conclusion that it is a neglected scientific discipline.

## 4. The origins of steganography

The inspiration of steganography is strongly related to phenomena observable in the animal and plant kingdoms. Evolution has proved long ago that impersonation is good protection and capacitates survival for numerous species. The ability to camouflage one's presence by means of adopting the characteristics of another living organism is referred to as mimicry. This capability permits certain organisms to improve their chances for survival. The Ancient Greeks, who sought inspiration in nature, had considered the ability to simulate its ways as a measure of craftsmanship. Inherently, the ancient people picked ordinary objects as a carrier for the secret message. The vector physical object, possibly even a living organism, had to be transported from one participant of communication to the other without raising suspicion on the way.

It should not be surprising that the first written report of the use of steganography is attributed to the Greek historian, Herodotus. The reported method involved camouflaging a secret message within a hare corpse [16]. The animal was meant to imitate a game trophy and was carried by a man disguised as a huntsman. In this way, a message could be passed without raising unnecessary suspicion.

The most notable method quoted in the historian's works is the communication on wooden tablets – these were usually coated with a thin layer of wax, on which text would be embossed. Clandestine passing of information with the aid of such medium could be achieved if the text was carved permanently on the wood (the carrier of the steganograms), and then coated with wax. Such object would then be passed as an unused tablet, and only an aware recipient would know that the letters would become visible if the wax coat was melted.

The Greek methods were fairly easy to implement as they relied on common patterns – the messages that were passed utilized a carrier cover that could be considered common at the time. Alongside the progress of human civilization and of the way people communicated, new opportunities arose. The popularization of parchment, which substituted papyrus, brought about a new cover for steganograms. Its popularity led to the development of complementary steganographic algorithms, capable of exploiting the new cover's properties. Pliny the Elder is considered the inventor of sympathetic inks [47], as he postulated the use of thithymallus plant's sap to write text, which would become invisible upon drying. A subtle heating process would lead to the charring of the organic substances contained in the ink, which would then turn brown.

The common factor of all of the aforementioned techniques is the operation of adding surplus content (additional features) to a carrier, which otherwise would not physically contain the inserted elements.

A different type of steganography that was invented in ancient Rome is the semagram, or a secret message which does not take written form. Tacitus, the historiographer of the ancient world, became interested in the Astragali [48], which were small dice made of bone. Such objects could be threaded onto a string, where the placement of the holes could be attributed meaning. A properly crafted object would pass unnoticed as a toy.

The Medieval Ages had brought about major progress in the art of information hiding. The Chinese invention of paper, upon its introduction to Europe, in the Middle Ages, had brought forth the necessity of differentiating between different manufacturers' products. This is how paper watermarking was born [43]. Presently, digital watermarking and digital image steganography, stemming from the mentioned invention, base on the same principle. It should be stressed that file watermarking is now considered a separate branch of the information hiding techniques. Petitcolas, Anderson and Kuhn, in their survey paper from 1999 [40], have derived a whole field of copyright marking, of which watermarking is a subclass. The current notion refrains from classifying digital watermarking as steganography, due to the lack of an explicit communication aspect and the inferior role of providing "invisibility" to the participants of communication and a larger importance of robustness of such embedded watermark.

The popularization of paper had further consequences. The steganographic vector was no longer necessarily a physical object, but could take written form, where the carrier text itself would conceal the privileged information. Among the inventions that achieved popularity during the Medieval times are the textual steganographic methods, among others, the acrostic. This term refers to pieces of writing, whose first letters or syllables spell out a message. The most famous example of such textual steganography is attributed to a Dominican priest named Francesco Colonna, who, in 1499 had hid in his book – "Hypnerotomachia Poliphili", a love confession which could be spelled out from the first letters of subsequent chapters [15].

More sublime carrier is the language itself, as the Medieval people had discovered. Here, the embedding process occurs in the linguistic syntax and semantics. Linguistic steganography may be derived from the aforementioned technique of textual steganography, as it relies on the manipulations on the written (possibly even spoken) language with the aim of tricking the perception of an unaware dupe.

Following the postulates of Richard Bergmair [10], linguistic steganography covers within its scope any technique which involves intentional mimicry of typical structures of words, characteristic to a specific language. This may concern the deliberate tampering with grammar, syntax and the semantics of a natural language. Any action involving modification of the aforementioned aspects should capacitate maintaining of the innocent appearance of the cover text.

The Renaissance had brought about an invention by an Italian-born scientist Giambattista della Porta who, in the XVI century, detailed how to hide a message inside a hard-boiled egg: write on the shell using ink made from a mixture of alum and vinegar. The solution penetrated the eggshell, leaving no trace on the surface, but a discoloration occurred on the white, leaving the message on its surface, which was only readable once the shell was removed.

Gaspar Schott, a German Jesuit from the Age of Enlightenment, followed the trail marked by his Renaissance predecessors. His work, published in 1680, entitled "Schola Steganographica", explained how to utilize music scores as a hidden data carrier. Each note corresponded to a letter, which appeared innocent as long as nobody attempted to play the odd-sounding melodies.

The Industrial Revolution, which followed the Age of Enlightenment, had brought about new means of communication. Newspapers became a popular and reliable source of the latest information. At some point it became obvious that a newspaper could serve as a perfect steganographic carrier. Since daily papers could be sent free of charge, it was convenient to poke holes over selected letters and thus craft a secret message. This is how 'newspaper codes' were born.

The first symptoms of the growing interest in steganography may be traced back to the period of the two World Wars and then the Cold War. These had brought about such steganographic techniques as microdots – punctuation marks with inserted microscopic negatives of images or texts [56].

The period of World Wars was a true bonanza of hidden communication schemes. First of the World Wars was witness to the spectacular return of all sorts of invisible inks [27]. The Second World War was marked by Hedy Lamarr's and George Antheil's patent for spread spectrum communication [34]. They devised a method for guiding torpedoes with a special, multi-frequency set of signals, resistant to jamming attempts. The control information was dispersed over a wide-frequency bandwidth, which provided cover. The idea of embedding information in a number of different frequencies later found use in the fields of digital image and audio steganography.

The technological development in the XX century had also accelerated the development of more sophisticated techniques. Among these inventions were the so-called "subliminal channels", which based on cryptographic ciphers for the embedding of steganograms. The main principle was to insert content into digital signatures. This concept was introduced in 1984 by Gustavus Simmons, despite the US government's prohibition on publishing of materials on steganography. Simmons proposed that the overt and monitored communication conducted between two participants can be supplemented with a steganographic channel. This channel would be based on a number of dedicated bits of the message authentication. These, at the cost of reducing the message authentication capability of the digital signature, would serve as the steganographic channel capacity [46]. The steganographic channel established in this way would be visible, yet undetectable. Subliminal channels utilized the cryptographic protocol as the carrier for steganograms.

## 5. Contemporary trends of development

Modern steganographic techniques utilize the XX[th] century's inventions – computers and networking. Four main trends of development of the so-called digital steganography can be distinguished:

- – digital media steganography,
- – linguistic steganography,
- – file system steganography,
- – network steganography.

These four main branches of digital steganography are explained and described below. It must be also emphasized that currently majority of research in that area is devoted to digital media and network steganography. The prior is a matured research area with significant achievements, thus the exploration of this field is presently not as dynamic as in the case of the recently sprouted group of techniques falling into the category of network steganography.

Digital media steganography dates back to the 1970's, when researchers focused on developing methods to secretly embed a signature in a digital picture. Many different methods were proposed, including, among others: patchwork, least significant bit modifications or texture block coding [8]. The introduced techniques were intended for both types of images: undergone lossy or lossless compression, like JPEG or BMP, which are the most common image formats. The variety of algorithms for the embedding in digital pictures can be grouped according to the type of alterations that were induced. Following Johnson and Jajodia, the modifications are either bit-wise – influencing the spatial domain characteristics of the image, or affect the frequency domain characteristics. Thirdly, specific file format intricacies may be exploited or, a mix of all these techniques is possible. The transform domain provides for the most versatile medium of embedding. Affecting of the image processing algorithms may involve, among others, discrete cosine transform (DCT), discrete wavelet transform (DWT), Fourier transform, which may result in alterations of e.g. luminance or other measurable property of an image [26, 19]. Digital image steganography's position is unfaltering – the survey paper by Cheddad et. al. [13] points that lately the interest concentrated on employing digital media steganography and watermarking for embedding of confidential, patient-related information in medical imagery. Another application of digital image steganography foreseen to become popular is the implantation of additional data in printed matter, which, invisible for the naked eye, becomes decodable, when photographed and processed by a cellular phone [13].

Notably, digital image steganography is mostly oriented towards tricking the human visual system into believing that the perception of the image has not been manipulated in any way [8]. Similar rule applies to the whole field of digital media steganography, whose primary function is to trick the observer to believe that the crafted "forgery" is indeed genuine. The communication aspect of the whole steganographic algorithm is secondary to the process of embedding of the secret data.

Alongside the development of digital image steganography, it appeared that the human auditory system is equally prone to delusion as the visual perception. The research focus moved to audio files like MPEGs. The developed techniques included, among others, frequency masking, echo hiding, phase coding, patchwork and spread spectrum. It also became apparent that error correction coding is a good supplemental carrier for audio steganography – any redundant data can be used to convey the steganogram at the cost of losing some robustness to random errors [8]. This idea later found use in network-protocol based steganography.

Next, steganographers took video files as target carrier. Most of the proposed methods were adaptations of the algorithms proposed for audio and image files. Video-specific solutions involved using either video's I-frames' color space [54] as a steganographic carrier or P-frames' and B-frames' motion vectors [58]. Currently, steganography in video files either takes advantage of the existing methods for audio and image files, or makes use of the intrinsic properties of the video transmission, like movement encoding.

Parallel to digital image and audio steganography, information hiding in text was developed – the available methods exploited various aspects of the written word. The first set of techniques altered word-spacing, which was even claimed to have been used at the times of Margaret Thatcher to track leakages of cabinet documents [5]. More advanced steganographic methods used syntactic and semantic structure of the text as a carrier. The introduced methods permitted for such displacement of punctuation marks, word order or alterations of the choice of synonyms, that could be attributed certain meaning. Presently, some suggest that even SPAM messages may be a carrier of steganography, due to the large amounts of such mail that is emitted every day [12]. According to work by Bennet [9], the possible techniques can either rely on the generation of text with a cohesive linguistic structure or use natural language text as a carrier. It should be pointed that the first technique does not fully fulfill the definition of steganography, where the existence of the carrier should be independent of the existence of the injected hidden content. Thus, a text lacking rhetorical structure, cannot be considered a proper carrier. Specialists also differentiate between textual steganography and linguistic steganography [9]. The "SPAM method" is a linguistic method, and the embedding occurs with the aid of Context Free Grammars. CFGs have a tree structure, thus the selection of proper words, or branches, provides encoding for binary data. An example of a textual method would be a substitution technique, where a message's carrier is the set of white-spaces and punctuations marks undergoing shifting, repetition or other modifications.

Parallel to this research, it was revealed that x86 machine code can also be subject to embedding [17]. Some amount of information can be placed in the carrier code, with the aid of careful selection of functionally-equivalent instructions. This method exploits the same principle as linguistic steganography, where the choice of words from the set of synonyms can be attributed steganographic meaning.

The invention of a steganographic file system by Anderson, Needham and Shamir was an eye-opener [6]. It became apparent that information can be steganographically embedded even in isolated computing environments. The main principle of steganogram preparation was similar to invisible inks – one that knew how to search, could reveal the encrypted files from a disk. The utilized mechanism relied on the fact that ciphered data resembles random bits naturally present on the disk and only the ability to extract the vectors marking the file boundaries permitted the location process. Another example of a steganographic file system can be found in [39], whose authors created a steganographic file system implementation on Linux. Their invention preserves the integrity of the stored files and employs a hiding scheme in the disk space with camouflaging with the aid of Dummy Hidden Files and Abandoned Blocks.

Alongside the abovementioned types of digital steganography, currently the target of increased interest is network steganography. This modern family of methods stems from "covert channels" – a number of techniques intended for monolithic systems, like mainframes. This term was first introduced by Lampson, who identified the problem of information leakage in non-confined programs [31]. The expression "network steganography" was coined by Szczypiorski [52]. Currently, the terms network steganography and covert channels are used interchangeably (and incorrectly), but historically they are sovereign of each other.

The summary of the evolution of the steganographic data carrier is presented in Fig. 1.

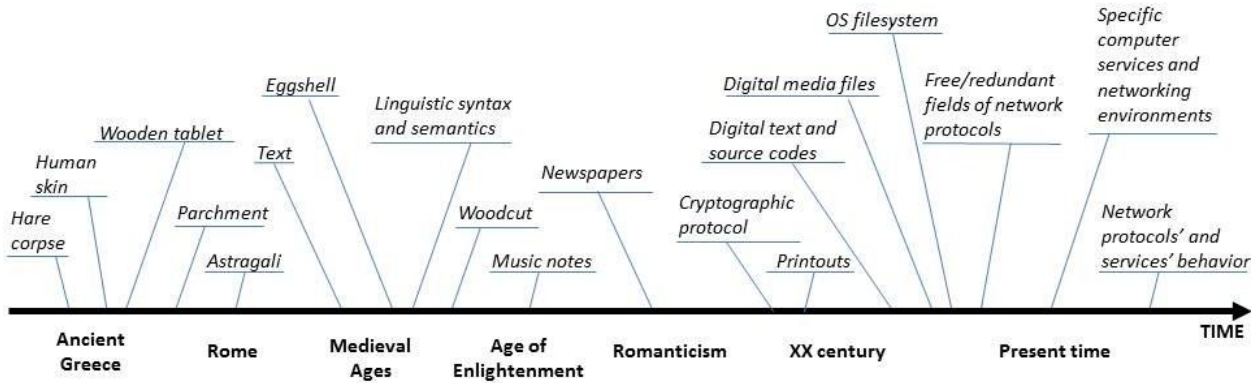

Fig. 1 Timeline of the evolution of hidden data carrier

## 5.1. Network Steganography: the youngest in the spotlight

Network steganography is the youngest branch of information hiding. It is a fast developing field: recent years have resulted in multiple new information hiding methods, which can be exploited in various types of networks. The exploitation of protocols belonging to the Open Systems Interconnection (OSI) reference model [59] is the essence of network steganography. This family of methods may utilize one or more protocols simultaneously or the relationships between them – relying on the modification of their intrinsic properties for the embedding of steganograms.

Network steganography is on the rise because embedding secret data into digital media files has been found to possess two serious drawbacks: it permits hiding only a limited amount of data per one file and the modified picture may be accessible for forensics experts (e.g. because it was uploaded to some kind of server). Network-level embedding changes the state of things diametrically; it allows for leakage of information (even very slow) during long periods of time and, if all the exchanged traffic is not captured, than there is nothing left for forensics experts to analyze. As a result, such methods are harder to detect and eliminate from networks.

Nowadays, network steganography relies on certain loopholes to conceal its presence. The first is the perceptual inability of the end user to sense minor differences between seemingly identical objects. For example, upon hearing a real-time audio recording transferred through a public network, a person almost certainly will not notice slight alterations of the transmitted voice, especially that he or she will lack any reference for that particular VoIP call. The second loophole permits the passage of steganograms through a network, without raising any alarms in the intermediate nodes. This typically relates to the statistical invisibility – that is, the induced anomalies do not exceed a reasonable threshold typical for network functioning. Typically, three characteristics of communications are utilized for steganographic purposes:

–   The communication channel is not perfect – errors are a natural phenomenon and thus it is possible to embed information in a pattern mimicking an ordinary distribution of damaged Protocol Data Units.
–   Most protocols bear some quantity of redundant information. The surplus fields can be used for embedding, if this does not induce malfunctioning of the carrier information flow.
–   Not every protocol is completely defined. Most of the specifications permit some amount of freedom in implementation, which can be abused.

Network steganography methods, following [25], can be broadly classified according to the number of protocols that are used for steganographic purposes. The modification of the properties of a single protocol from the OSI model is called intra-protocol steganography, whereas exploitation of relationships between multiple protocols is classified as inter-protocol steganography. Once a protocol or number of protocols are chosen as a carrier for secret data, it is decided how the embedding should be performed. The first possibility is to inject the covert information into the Protocol Data Unit (PDU) – this can be done by means of modification of protocol specific fields or by means of insertion into the payload, or both. Alternatively, or complimentary to the previous technique, it is possible to modify the time relations between the PDUs. These changes may impact the order of PDUs, their losses or their relative delays. Hybrid methods utilize both – modification of PDUs and their time relations.

The predecessor of current, more sophisticated network steganography methods, was the utilization of different fields of TCP/IP stack's protocols [42] as a hidden data carrier. Majority of the early methods concentrated on the embedding in the unused or reserved fields of protocols to convey secret data. Then, more advanced methods were invented, which were targeted towards specific environments or towards specific services. Recent solutions exploit:

–   Multimedia, real-time services like IP telephony [33],
–   Popular Peer-to-Peer services: Skype [35] or P2P file sharing systems like BitTorrent [30],
–   Social media sites like Facebook [37]
–   Wireless network environments: e.g. Wireless Local Area Networks (WLAN) [53], or Long Term Evolution (LTE) [22],
–   Cloud computing environment [41],
–   New network protocols, like SCTP [20] .

Although the use of IP telephony service as a hidden data carrier can be considered a fairly recent discovery [33], the existing VoIP steganographic methods stem from two distinct research origins. The first is the aforementioned, well-established digital media steganography, which has given rise to methods that target the digital representation of the transmitted voice as the carrier for hidden data. The second sphere of solutions target specific VoIP protocol (e.g. signaling, transport or control protocols) fields, or the protocol's behavior.

Lately, another steganographic method – TranSteg (Transcoding Steganography), intended for a broad class of multimedia and real-time applications, like IP telephony, has been proposed [36]. TranSteg bases on the general idea of transcoding (lossy compression) of the voice data from a higher bit rate codec, and thus greater voice payload size, to a lower bit rate codec with smaller voice payload size. This occurs with the least possible degradation in voice quality. In other words, compression of the overt data is utilized to make space for the steganogram in the payload field. The achieved steganographic bandwidth is as high as 32 kbit/s.

Looking into the P2P services' steganographic applicability, one may encounter a steganographic method named SkyDe (Skype Hide), proposed for Skype by Mazurczyk et al. [35]. It utilizes encrypted Skype voice packets as a hidden data carrier. By taking advantage of the high correlation between speech activity and packet size, packets without voice signal can be identified and used to carry secret data. This is achieved by replacing the encrypted silence with secret data bits. The resulting steganographic bandwidth, or hidden-data rate (amount of secret data that can be sent per unit of time, when using a particular method) is about 2 kbit/s.

Another recent invention for an Internet P2P service, the StegTorrent, has been introduced for the BitTorrent application [30]. StegTorrent takes advantage of the fact that, in BitTorrent there are usually many-to-one transmissions, and that for one of its specific protocols – μTP – the header provides means for packets' numbering and retrieving their original sequence. This allows for sending hidden data with a rate of about 270 b/s.

For social media sites like Facebook – the authors of [37] proposed creating a botnet communicating over unobservable communication channels. The bots exchanged information with their botmaster by embedding information in images and using the image sharing capabilities to route the secret data to the recipient.

When it comes to the "wireless environment", different standards are targeted by steganographers. For example, for WLANs, Szczypiorski and Mazurczyk have introduced a method called WiPad (Wireless Padding) [53]. The technique is based on the insertion of hidden data into the padding of frames at the physical layer of WLANs. It allows to pass data in a covert way with a significantly high data rate of about 1.5 Mbit/s. A similar concept was utilized in [22] for LTE (Long Term Evolution) and the resulting data rate was about 1.2 Mbit/s.

The cloud computing environment, which, in the eyes of Ristenpart et al. is vulnerable to cross-Virtual Machine information leakage [41], is a great playground for exercising steganography. They proposed a range of techniques for obtaining classified information by probing the values of shared-cache load, CPU load, keystroke activity, or similar.

Other promising future-network protocols, like, for example, the SCTP (Stream Control Transmission Protocol) which is a candidate for new transport layer protocol, and might replace TCP (Transmission Control Protocol) and UDP (User Datagram Protocol) protocols, is also prone to steganography. Detailed analysis in [20] reveals the most likely 'places' in SCTP transmissions that could be utilized for information hiding. Special attention is directed towards steganographic methods that utilize new features, characteristic to SCTP, such as multi-homing and multi-streaming.

To summarize, various network services and applications can and will become target of embedding, and the larger the proliferation of a certain service or application, the more attractive it is to piggyback secret data by means of network steganography.

# 6. Conclusions

Information hiding covers within its scope various techniques intended for the communication of messages with the aim of keeping some aspect of such exchange secret. This may involve providing security by obscurity for the participants of the dialogue (anonymity), secrecy of the messages (steganography) or protection of the carrier (copyright marking).

The roots of the described methods stem from antique times. The need for sending messages which cannot be compromised in case of interception had motivated people to create codes or symbols which appeared innocent, but in fact had different significance than the apparent.

Modern information hiding employs various embedding techniques – a lot of these are the result of the transfer of some previously known method into the digital domain. An interesting exception to this notion is network steganography, a family of methods which emerged with the popularization of networked environments. The appearance of new secret data carriers in steganography can be treated as evolutionary steps in the information hiding techniques' development. The growing number of communication protocols, services and computing environments offers almost unlimited opportunities for displaying a whole spectrum of steganographic methods. It is noteworthy that it is the carrier's properties as well as its popularity that predestine or limit its capability to serve as an efficient medium for clandestine communication, and the emergence of a new technology will likely bring about new information embedding opportunities.

Illicit activities conducted in the virtual world pose a tangible threat to the society, as recent cyber warfare events show. Indisputably, information hiding has joined the arsenal of the utilized weapons, and thus it should be recognized that it poses a large threat to information systems' security. More importantly, the matter is pressing, because steganalysis techniques are still one step behind the newest steganography methods. There is no "one size fits all" solution available and ready to detect covert communication in our current network security defenses systems. Thus, we urge the research community to focus its efforts to promptly come up with steganalysis methods that could be practically deployed in networking environments.